%% file: beehive-asplos-17.tex
\documentclass[sigconf, 10pt]{acmart}
\renewcommand\footnotetextcopyrightpermission[1]{} % removes footnote with conference information in first column
\setcopyright{none}

\usepackage[normalem]{ulem}
\usepackage{caption}
\usepackage{subcaption}
\usepackage{listings}
\usepackage{relsize}
\usepackage{url}
\usepackage{xcolor}

\usepackage{caption}
\usepackage{subcaption}
\usepackage{listings}
\usepackage{relsize}
\usepackage{url}
\usepackage{xcolor}
\usepackage{graphicx}
\usepackage{multirow}
\usepackage{array}
\usepackage{booktabs}

\graphicspath{{./images/}}
\DeclareGraphicsExtensions{.pdf}

\lstdefinestyle{beehive}{
language=Java,
morekeywords={@Parallel,@Read,@Write,@ReadWrite}
}

\lstdefinestyle{cuda}{
	language=C++,
	morekeywords={__global__}
}

\begin{document}

\title{Project Beehive: A Hardware/Software Co-Designed Platform for Runtime and Architectural Research}
\author{Christos Kotselidis, Andrey Rodchenko, Colin Barrett, Andy Nisbet, John Mawer, Will Toms, James Clarkson, Cosmin Gorgovan, Amanieu d'Antras, Yaman Cakmakci, Thanos Stratikopoulos, Sebastian Werner, Jim Garside, Javier Navaridas, Antoniu Pop, John Goodacre, Mikel Luj\'an}
\affiliation{Advanced Processor Technologies Group\\The University of Manchester}
\affiliation{first.last@manchester.ac.uk}
%r\author{The University of Manchester}
% \affil{Advanced Processor Technologies Group\\The University of Manchester}
 %r \author{first.last@manchester.ac.uk}

%\author{Advanced Processor Technologies Group\\
%	The University of Manchester}
\date{}
\maketitle

\input{abstract}

\input{introduction}

\input{architecture}

\input{usecase}

\input{results}

\input{related_work}

\input{conclusions}

%\acks
%The research leading to these results has received funding from UK EPSRC grants DOME EP/J016330/1, AnyScale Apps  EP/L000725/1, INPUT EP/K015699/1 and PAMELA EP/K008730/1 and the European Union's Seventh Framework Programme under grant agreement n° 318633 AXLE project, and  n° 619788 RETHINK big. Mikel Lujan is funded by a Royal Society University Research Fellowship and Antoniu Pop a Royal Academy of Engineering Research Fellowship.

\section{Acknowledgements}
The research leading to these results has received funding from UK EPSRC grants DOME EP/J016330/1, AnyScale Apps  EP/L000725/1, INPUT EP/K015699/1 and PAMELA EP/K008730/1 and the EU FP7 Programme under grant agreement No 318633 AXLE project, and  EU H2020 No 732366 ACTiCLOUD. Mikel Lujan is funded by a Royal Society University Research Fellowship and Antoniu Pop a Royal Academy of Engineering Research Fellowship.

\bibliographystyle{plain}
\bibliography{paper}

\end{document}

%% file: abstract.tex
\begin{abstract} 	
With the extreme scaling of current architectures, from wearables to exascale systems, along with new application domains such as Big Data and human-centred applications, vertical and cross-cutting research is vital.
Solutions based solely in hardware or software are no longer sufficient to meet the requirements of today's ubiquitous computing or maintain the pace of improvements seen during the past few decades.

In hardware, the well cited end of single-core scaling has resulted in the proliferation of multi-core system architectures forcing complex parallel programming techniques into the mainstream. 
To further affect the exploitation of physical resources, systems are becoming increasingly heterogeneous with specialized computing elements and accelerators. 
Programming across such a range of disparate architectures requires a new level of abstraction and adaptation by programming languages and applications. 

In software, for example, emerging complex applications from domains such as Big Data and Computer Vision, run on multi-layered software stacks targeting hardware with a variety of constraints and resources. 
The design space of current and future computing is becoming extremely broad making the optimization task challenging.
Multi-objective optimization for power, performance, and resiliency requires experimentation platforms facilitating quick and easy prototyping of hardware and software with intimate co-designed techniques. 

In this paper, we present Beehive: A Hardware/Software co-designed platform enabling simultaneous runtime and architectural research.
Beehive utilizes various state-of-the-art software and hardware components along with novel and extensible co-designed tools and techniques. 
The objective of Beehive is to provide a flexible platform for rapid prototyping and experimentation across the emergent range of applications, programming languages, compilers, runtimes, and low-power heterogeneous many-core architectures in a full-system co-designed manner.
We use a complex Computer Vision application, as a use-case, to showcase the versatility and effectiveness of Beehive by accelerating it across numerous and diverse metrics achieving up to 43x performance improvements.
\end{abstract}

%% file: introduction.tex
\section{Introduction}
\label{introduction}

Traditionally, software and hardware providers have been delivering significant performance improvements on a yearly basis. 
Unfortunately, this is no longer feasible.
Predictions about "dark silicon" \cite{Esmaeilzadeh:2011:DSE:2000064.2000108} and resiliency \cite{Shafique:2014:ECD:2593069.2593229}, especially in the forthcoming exascale era \cite{Cappello}, suggest that traditional approaches to computing problems are impeded by power constraints and process manufacturing.
Furthermore, since single-threaded performance has been saturated both at the hardware and the software layers, new ways for pushing the boundaries have emerged.
After the introduction of multi and many core systems, heterogeneous computing and ad-hoc acceleration, via ASICs and FPGAs \cite{stella, zynq}, are advancing into mainstream computing.

The extreme scaling of current architectures, from low-power wearables to high-performance computing, along with the diversity of programming languages and software stacks, create a wide spectrum of space exploration for achieving optimal energy-efficient results. 
Co-designing an architectural solution at the system-level\footnote{In this context we refer to architectural solution as a co-designed solution that spans from a running application to the underlying hardware architecture.} requires tight integration and collaboration between teams, that have typically been working in isolation. 
The design-space to be explored is vast, and there is the potential that a poor, even if well intentioned, decision will propagate through the entire co-designed stack.
Thus, amending the consequences at a later date may prove extremely complex and expensive, if not impossible.

In this paper we present Beehive: a complete full-system hardware/software co-designed platform for rapid prototyping and experimentation (All the available hardware and software components of Beehive will be publicly available).
Beehive enables co-designed optimizations from the application level down to the system and hardware level, enabling accurate decision making for architectural and runtime optimizations.
As a use-case, we accelerate and optimize the complex KinectFusion \cite{newcombe;2011;kinectfusion:-r} Computer Vision application in numerous ways through Beehive's highly integrated stack achieving up to 43x performance improvements.

In detail, Beehive makes the following contributions:
\begin{itemize}
\item \textbf{Enables co-designed research and development for traditional and emerging applications and workloads}: To achieve this, we tightly integrate the software and hardware layers of the stack in a unified manner while expanding Beehive's reach to complex applications and workloads (Section \ref{Applications}). We showcase that capability by implementing a Java-based version of KinectFusion and co-designing it through Beehive's stack.
\item \textbf{Enables co-designed compiler and runtime research for multiple dynamic and non-dynamic programming languages in a unified manner}: This is achieved by unifying under the same compilers and runtimes, high-quality production and research Virtual Machines able to execute transparently multiple programming languages (Section \ref{max}).
\item \textbf{Enables heterogeneous processing on a variety of platforms such as ARM (ARMv7 and Aarch64), and x86}: The unified runtime layer has been extended to support multiple ISAs scaling from high-performing x86 to low-power ARM architectures (Section \ref{Runtime}). We showcase that capability by evaluating standard benchmarks along with the KinectFusion use case.
\item \textbf{Provides fast prototyping and experimentation on heterogeneous programming on GPGPUs, SIMD units, and FPGAs}: The novel Tornado, Indigo, and MAST modules achieve transparent heterogeneous execution on GPGPUs, SIMD units, and FPGAs respectively, without sacrificing productivity (Sections \ref{tornado}, \ref{indigo}, \ref{mast}). We showcase that capability by accelerating KinectFusion on GPGPUs, SIMD units, and FPGAs under the same infrastructure.
\item \textbf{Enables co-designed architectural research on power, performance, and resiliency techniques via high-performing simulators and real hardware}: Along with a plethora of real hardware, Beehive integrates a number of high-performing simulators in a unified framework (Section \ref{simul}). We showcase this capability by providing a novel hardware/software co-designed optimization for KinectFusion.
\item \textbf{Supports dynamic binary optimization techniques via instrumentation and optimization at the system and chip level}: Beehive extends its research capabilities to novel micro-architectures by providing dynamic binary instrumentation and optimization techniques for all supported hardware architectures (Section \ref{bininst}).
\end{itemize}

The paper is organized as follows: Section \ref{architecture} explains the architecture of Beehive along with its individual components.
Section \ref{usecase} presents the Computer Vision application that forms the use case in this paper.
Section \ref{results} presents the various co-designed optimizations applied to the selected application along with their correspondent performance evaluations. 
Finally, Sections \ref{related} and \ref{conclusions} present the related work, the concluding remarks and the future vision of Beehive, respectively.

%% file: architecture.tex
\section{Beehive Architecture}
\label{architecture}

\subsection{Overview}

\begin{figure}
\centering
\includegraphics[height=94mm, width=\columnwidth]{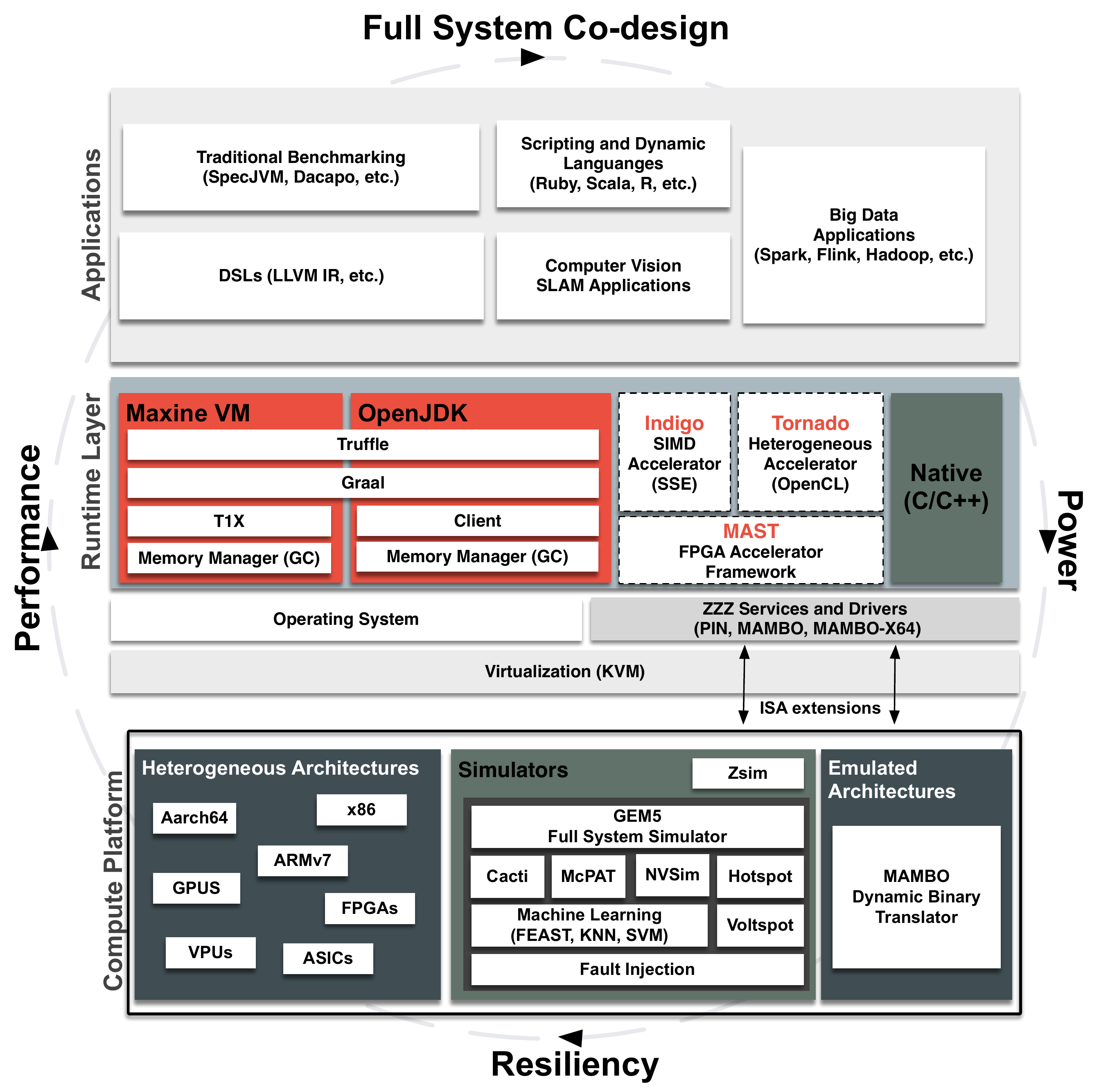}
\caption{Beehive architecture overview.}
\label{beehive_stack}
\end{figure}

Beehive, as depicted in Figure \ref{beehive_stack}, follows a multi-layered approach of highly co-designed components spanning from the application down to the hardware level.
The design philosophy of Beehive revolves around five pillars: 
\begin{enumerate}
	\item \textbf{Rapid prototyping} for developing full-stack optimizations efficiently by using high-level programming languages.
	\item \textbf{Diversity} for tackling multiple application domains, programming languages, and runtime systems in a unified manner.
	\item \textbf{Accuracy} of obtained results by integrating and augmenting state-of-the-art industrial-strength components.
	\item \textbf{Maintainability} of the platform keeping it on par with the state-of-the-art in the long term. 
	\item \textbf{Scalability} of the platform to complex systems and architectures in a seamless manner.
\end{enumerate}

Beehive targets a variety of workloads ranging from traditional benchmarks to emerging applications from a variety of domains such as Computer Vision and Big Data.
Furthermore, as explained later in Section \ref{Applications}, Beehive allows multiple implementations of complex applications in a variety of programming languages in order to enable comparative research amongst them.
Beehive supports both managed and un-managed languages as explained in Subsection \ref{Runtime}.
Finally, applications can execute either directly on hardware, in-directly on hardware using a dynamic binary optimization layer, or inside Beehive's simulator stack.

The following subsections explain in detail each layer of Beehive along with the supported applications, programming languages, and hardware platforms.
 
\subsection{Applications}
\label{Applications}
Beehive targets a variety of applications in order to enable co-designed optimizations in numerous domains. 
Whilst compiler and micro-architectural research traditionally uses benchmarks such as SpecCPU \cite{speccpu}, SpecJVM \cite{specjvm}, Dacapo \cite{DaCapo:paper}, PARSEC \cite{bienia11benchmarking}, Beehive also considers complex emerging application areas. 
The two primary domains targeted by Beehive are Computer Vision applications and algorithms such as KinectFusion \cite{newcombe;2011;kinectfusion:-r} and other SLAMs (Simultaneous Localization and Mapping algorithms) along with Big Data software stacks such as Spark \cite{;;apache-spark}, Flink \cite{apache-flink}, and Hadoop \cite{;;apache-hadoop}.
To showcase Beehive, we selected an implementation of KinectFusion to be the main vehicle of experimentation.

Recent advances in real-time 3D scene understanding capabilities can radically change the way robots interact with and manipulate the world.
A proliferation of applications and algorithms, in recent years, have targeted real time 3D space reconstruction both in desktop and mobile environments \cite{DysonLab, ProjectTango,  newcombe;2011;kinectfusion:-r}.
To assess both the accuracy and performance of the proposed optimizations, we use SLAMBench \cite{2015PAMELASLAMBench} a benchmarking suite that provides a KinectFusion implementation.
SLAMBench harnesses the ICL-NUIM dataset \cite{2014Handa} of synthetic RGB-D sequences with trajectory and scene ground truth for reliable accuracy comparison of different implementations and algorithms. 
SLAMBench currently includes implementations in C++, CUDA, OpenCL, and OpenMP allowing a broad range of languages, platforms, and techniques to be investigated.
In Section \ref{usecase}, SLAMBench is explained and decomposed to its key kernels.

\captionsetup[figure]{skip=0pt}
\begin{figure*}
\includegraphics[width=\textwidth]{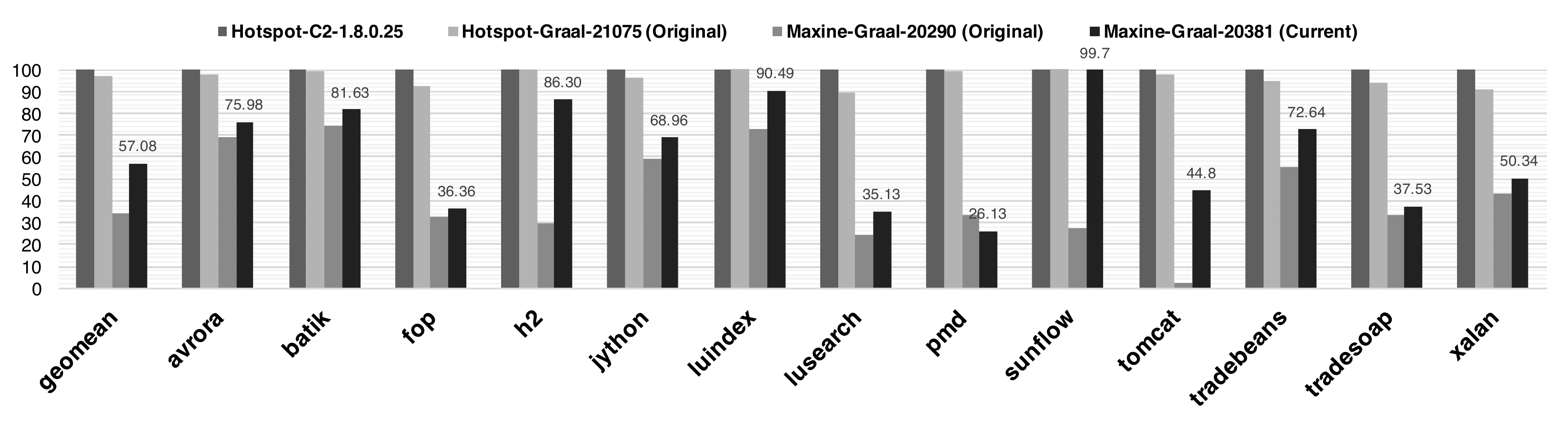}
\caption{DaCapo-9.12-bach benchmarks (higher is better) normalized to Hotspot-C2-Original.}
\label{figure:DaCapoResults}
\vspace{1mm}
\includegraphics[scale=0.7, width=\textwidth]{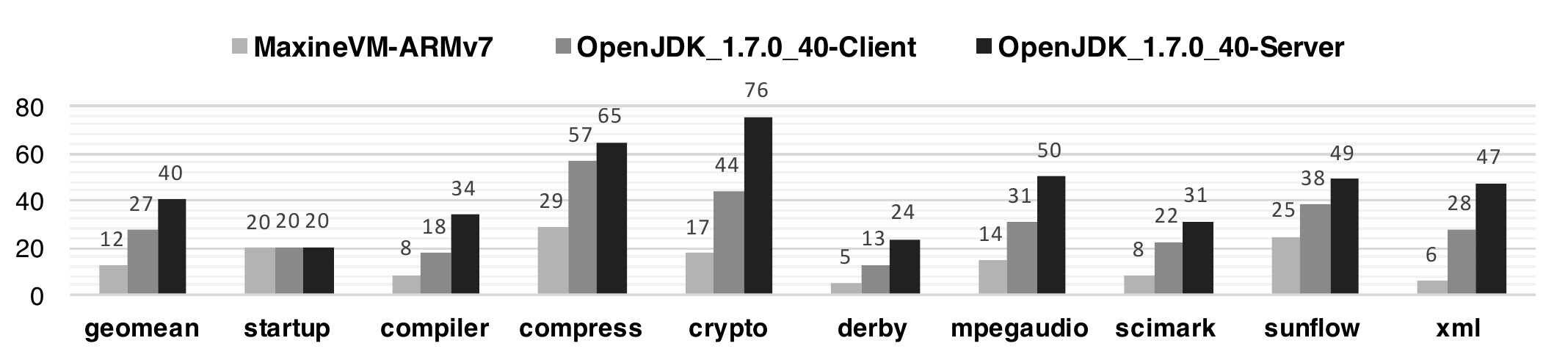}
\caption{SpecJVM2008 benchmarks (higher is better) normalized to OpenJDK-Zero-IcedTea6\_1.13.11.}
\label{figure:SpecResults}
\end{figure*}

\subsection{Runtime Layer}
\label{Runtime}
Some of the key features of Beehive are found in its runtime layer, which provides capability beyond simply running native applications.
One of the challenges when designing such tightly co-designed systems is the application and programming languages support.
Supporting numerous runtimes with various back-ends and compilers, while seamlessly integrating them with the lower layers of the computing stack, is a time consuming task which impedes
the maintainability of the whole platform.
These issues in turn will manifest in slow adoption of state-of-the-art software and hardware components and applications.

In order to overcome these challenges, we have taken the design decision to build the runtime layer around two components: the Java Virtual Machine (JVM) and native C/C++ applications.
Despite being able to execute native C/C++ applications (regardless of the compiler used), Beehive has been designed to target languages that can run, and be optimized, on top of the JVM.
The advent of the Graal compiler \cite{duboscq;2013;graal-ir:-an-ex} along with the Truffle AST interpreter \cite{Wurthinger:2013} enables the execution of multiple existing \footnote{For example, languages such as  Ruby, JavaScript, R, LLVM-based, etc., are currently supported by Truffle.}, and novel, dynamic and non-dynamic programming languages and DSLs on top of the JVM.
Building the Beehive platform around Truffle, Graal, and the JVM, we achieve high performing execution of a variety of programming languages in a unified manner.
Furthermore, the amount of maintenance necessitated is contained to two compilers and one runtime system.
In addition, any changes from the open sourced Graal and Truffle projects can be down-streamed to Beehive; keeping it synchronized with the latest software components.

Regarding the runtime systems of Graal and Truffle, two design alternatives have been deployed.
The first route is the vanilla implementations running on top of OpenJDK.
The benefits of this approach is that Beehive can be utilized by industrial-strength, high-performing systems that run on top of OpenJDK.
This, however, has a number of drawbacks.
Components of the runtime layer such as Object Layouts, Garbage Collection (GC) algorithms, Monitor Schemes, etc., are difficult to research due to the lack of modularity in OpenJDK.
To that end, we decided to add an additional runtime layer for Graal and Truffle: the Maxine Research Virtual Machine \cite{Wimmer}.

The MaxineVM, a meta-circular Java-in-Java VM developed by Oracle Labs, has been adopted and augmented for usage in Beehive \cite{kotselidis;2017;heterogeneous-m}.
Since its last release from Oracle, it has been enhanced by the Beehive team both in performance and functionality terms (Section \ref{max}).
The Graal compiler ported on top of MaxineVM has been stabilized and its performance has been improved making MaxineVM the highest performing research VM (Section \ref{max}).
In addition, as depicted in Figure \ref{beehive_stack}, both MaxineVM and OpenJDK use the same optimizing compiler accompanied by the Truffle AST interpreter enabling Beehive to extend its research capabilities from industrial strength to high-quality research projects.

The multi-language capabilities of Beehive have been further augmented by novel software components that enable heterogeneous execution of applications on numerous hardware devices; Indigo, Tornado, and MAST \cite{mast, kotselidis;2017;heterogeneous-m}.
While Indigo enables the exploitation of SIMD units, Tornado targets GPGPUs and FPGAs by OpenCL code emission.
Furthermore, MAST provides a clean API to access FPGA modules in a concurrent and thread-safe manner.
The following subsections explain in detail MaxineVM, Indigo, Tornado, while MAST is explained in Section \ref{mast}.

\subsubsection{MaxineVM}~
\label{max}

\noindent The latest release of MaxineVM from Oracle had the following three compilers:
\begin{enumerate}
	\item T1X: A fast template-based interpreter (stable).
	\item C1X: An optimizing SSA-based JIT compiler (stable).
	\item Graal: An aggressively optimizing SSA-based JIT compiler scheduled to be integrated in OpenJDK Java9 (semi-stable).
\end{enumerate}

\noindent Furthermore, MaxineVM was tied to the x86\_64 architecture. 
In the context of Beehive the following enhancements has been made to MaxineVM:
\begin{enumerate}
	\item T1X: Added profiling instrumentation enabling more aggressive profile-guided optimizations.
	\item T1X: Compiler ports to ARMv7 and Aarch64 enabling experimentation on low-power 32bit and 64bit architectures.
	\item C1X: Compiler port to ARMv7 enabling experimentation on low-power ARM 32bit architectures.
	\item Graal: Stability and performance improvements.
	\item Maxine: Complete ARMv7 and undergoing Aarch64 support, stability, and performance enhancements.
\end{enumerate}

\noindent Figures \ref{figure:DaCapoResults} and \ref{figure:SpecResults} illustrate the performance of MaxineVM in x86 and ARMv7 on Dacapo9.12-bach \cite{DaCapo:paper} and SpecJVM2008 \cite{specjvm} respectively.

As illustrated in Figure \ref{figure:DaCapoResults}\footnote{Intel(R) Core(TM) i7-4770@3.4GHz, 16GB RAM, Ubuntu 3.13.0-48-generic, 16 iterations, 12GB heap.}, since Oracle's last release (Maxine-Graal-rev.20290 Original), performance has been increased by 64\% (Maxine-Graal-rev.20381 Current) while currently Maxine is half of the performance of industrial strength OpenJDK with the C2 and Graal (rev. 21075) compilers.
The target is to get the JIT performance of both VMs on par by enabling more aggressive Graal optimizations in Maxine such as escape analysis \cite{stadler;2014;partial-escape-} and other compiler intrinsics.
Unfortunately, we could not compare against JikesRVM \cite{Alpern:2000:JVM:1011388.1011400} since it can not run the Dacapo9.12-bach benchmarks on x86\_64.

Regarding ARMv7, as depicted in Figure \ref{figure:SpecResults}\footnote{Samsung Chromebook, Exynos 5 Dual@1.7GHz, 2GB RAM, Ubuntu 3.8.11, 2GB heap.} the performance of MaxineVM-ARMv7 falls between the performance of OpenJDK-Zero and OpenJDK-1.7.0-(Client, Server).
MaxineVM outperforms OpenJDK-Zero by 12x on average across SpecJVM2008\footnote{\texttt{Serial} was excluded from the evaluation.}, while it is around 0.5x and 0.3x slower than the OpenJDK-1.7.0 client and server compilers respectively. 
As in x86, many optimizations both in the compiler and the code generator, will be implemented and/or enabled in order to match the performance of the industrial strength OpenJDK.

Regarding the memory manager (GC), various options are being explored ranging from enhancing Maxine VM's current GC algorithms to porting existing state-of-the-art memory management components. 
Currently MaxineVM supports semi-space and generational schemes.

\subsubsection{Indigo}~
\label{indigo}

\begin{figure}
\centering
\includegraphics[width=80mm]{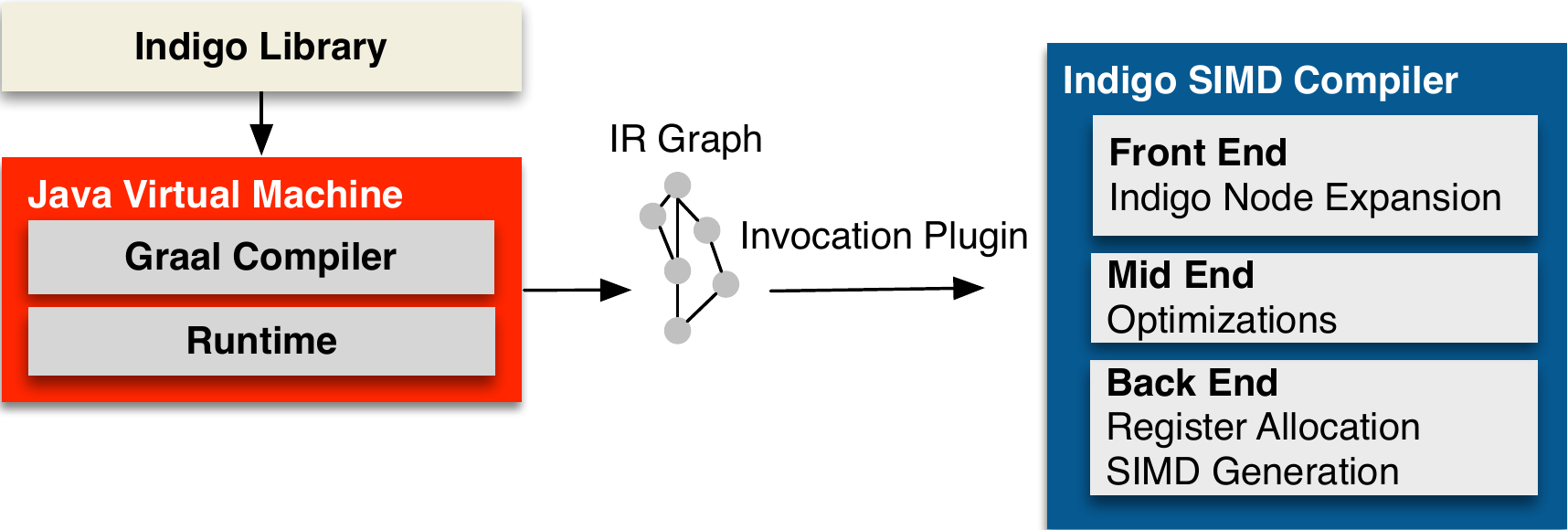}
\caption{Indigo's interaction with the Graal compiler.}
\label{fig:indigo-outline}
\end{figure}

\noindent Indigo, a novel component of Beehive, is an extension plugin for Graal that provides efficient execution of short vector types, commonly found in Computer Vision applications, and support for SIMD execution. 
While Indigo was initially designed to enhance the performance of computer vision applications, it can be easily expanded to provide generic vectorization support in Graal; a feature which is currently missing from public distribution.
Figure \ref{fig:indigo-outline} outlines how Indigo operates with the Graal compiler.

As depicted in Figure \ref{fig:indigo-outline}, Indigo uses Graal's invocation plugin which enables the custom addition of a node in Graal's Intermediate Representation (IR).
This, in turn, can be exploited by Indigo to re-direct the compilation route from Graal to Indigo and use its compilation stack to compile and optimize for SIMD execution.
Within Graal, the IR is maintained as a structured graph with nodes representing actions or values while edges represent their dependencies.
The graph is initially generated by parsing the bytecode from a class file.

The objective of vectorization is to reduce the distance between vector operations in the IR enabling further optimizations through virtualization (i.e. escape analysis and scalar replacement \cite{stadler;2014;partial-escape-}).
With the use of virtualization, we can maintain temporary vectors entirely at the registers of the targeted architectures.
The addresses of the vectors are being used for reading and writing, enabling us to break free from the primitive Java types and, more importantly, from the use of Java arrays.
However, since this is not an inherent safe usage of the Java semantics we made the following assumptions:

\begin{itemize}
\item Hardware supports 128-bit vector operations, true for ARM NEON and Intel SSE implementations.
\item The class contains four single-precision floating point numbers suitable for vector operations of SLAM applications.
\item Unused elements of a vector are zero.
\item The elements of a vector are contiguous in memory.
\item Once constructed, a vector is immutable.
\end{itemize}

The aforementioned assumptions apply to the library provided by Indigo and in turn allow some of the restrictions in Java to be eliminated.
This enables the IR to be extended and optimized more aggressively since the semantics are now within the vector abstraction and not within the general purpose language.

Invocation plugins allow the replacement of a method invocation with a sub-graph created during the graph building phase in Graal.
We used a single node plugin that contains its own domain specific compiler stack.
The major benefit of this approach is the runtime independence from Graal.
Therefore, it can be downloaded and used a standalone library that, if the JVM uses Graal on top of the JVM Compiler Interface (JMVCI) \cite{JEP243}, SIMD instruction emission can be generated.
Indigo's compiler stack contains a basic graph builder, optimizer, register allocator, and code generator with a scope limited for its target domain: Computer Vision applications.

Indigo nodes are generated either during the graph building phase of the compilation or indirectly during inlining.
Once a graph has been constructed, it is transformed during the optimization phases by exploiting canonicalization and simplification to merge nodes.
This allows us to maximize the number of operations in the node and eliminate \texttt{new instance} nodes (allocation of new objects) from the graph, leaving the data in registers.
A simplification phase traverses the operand edges of the Indigo node to detect other Indigo nodes and merges the internal operation graphs together.

When Indigo nodes are lowered to the low-level IR (LIR) nodes used by Graal, they must claim virtual registers from Graal.
At this point we lower the operation to a generic SIMD instruction to be scheduled while profiling the register requirements.
In order to maintain the vanilla implementation of Graal, we indirectly use its register allocator to provide general purpose and vector registers by claiming values to satisfy the requirements of the compiled method.
Later, these will be converted into physical registers during the back end phases.
The use of profiling enables us to offload the allocation algorithms to Graal, while ensuring that no vector registers are spilled to the stack.
This technique prohibits the JVM from entering un-recoverable states while being spatially more efficient.

Thanks to the modularity of Graal, and access to the compiler through the JVMCI, it is possible to insert novel nodes into the compiler at runtime.
With Indigo we show that it is possible to add a domain specific compilation plugin to augment the Graal compiler.
This allows us to bypass all Graal internals and emit machine code exploiting SIMD instructions that are unsupported in the publicly available Graal.
While this approach targets idiomatic SIMD for Computer Vision, there is no technical reason why it cannot be extended to insert other domain specific knowledge into Java.

\begin{figure}
\centering
\includegraphics[width=\columnwidth]{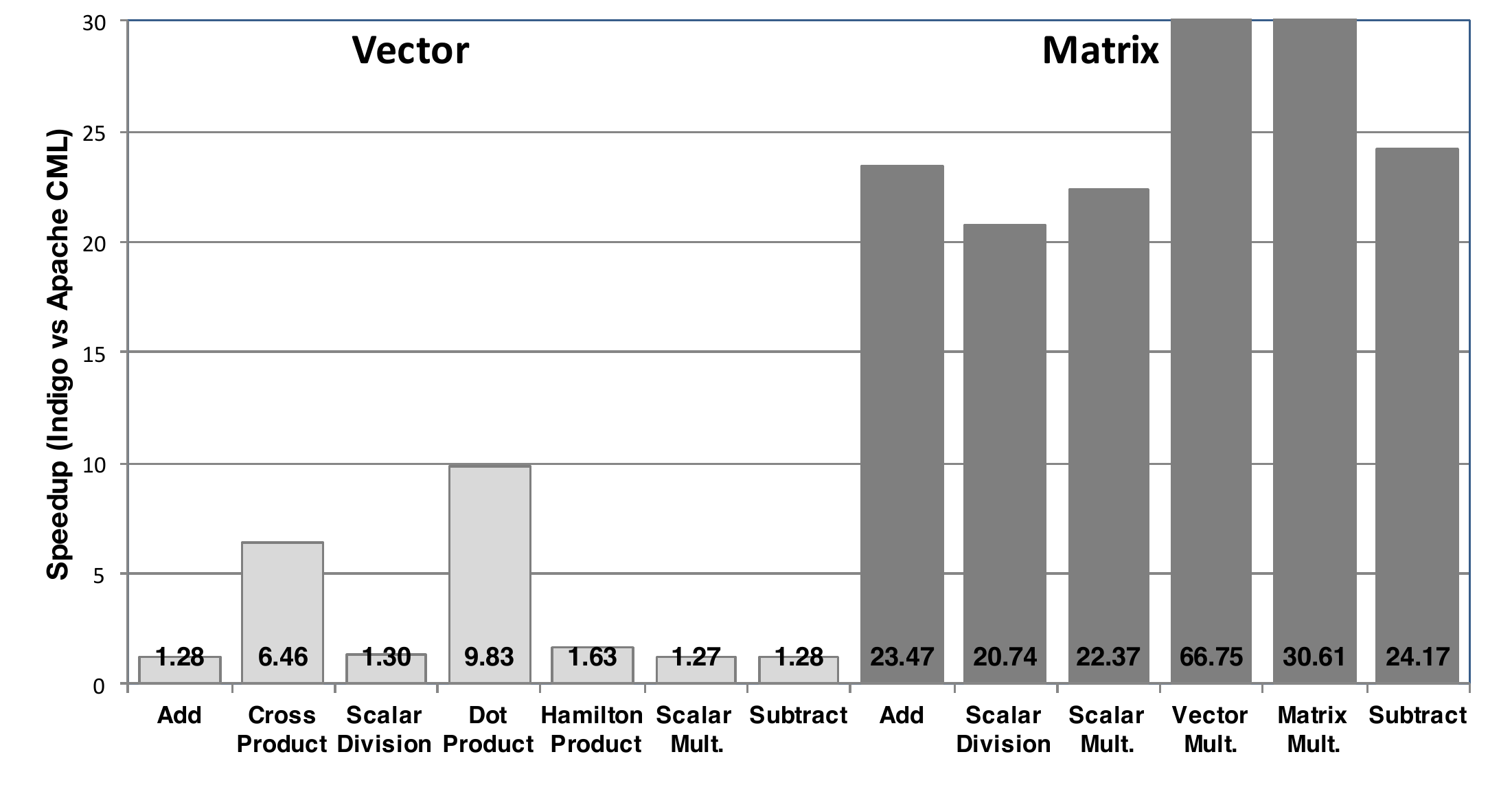}
\caption{Indigo's performance against Apache CML on common vector and matrix operations.}
\label{fig:indigo-perf}
\end{figure}

Figure \ref{fig:indigo-perf}, contains Indigo's relative performance against the Apache Common Mathematics Library (CML) \cite{Apache:2016} for a total number of 13 vector and matrix operations commonly found in Computer Vision applications.
As depicted in Figure \ref{fig:indigo-perf}, Indigo outperforms Apache CML both in vector and matrix operations.
As expected, the largest gains are observed in matrix operations with matrix-vector multiplication exhibiting a 66.75x speedup.
The observed performance improvements derive from the use of SIMD execution along with the compiler optimizations provided by Indigo (null check elimination, scalar replacement, etc.).

\subsubsection{Tornado}~
\label{tornado}

\noindent Tornado, a novel component of Beehive, originated by JACC \cite{clarkson2017boosting}, is a framework designed to improve the productivity of developers targeting heterogeneous hardware.
By exploiting the available heterogeneous resources, they have the potential to improve the performance and energy-efficiency of their applications.
The key difference between Tornado and existing programming languages and frameworks is its \textit{dynamism}; developers do not need to make a priori decisions about their hardware targets.
The Tornado runtime system achieves transparent computation offloading with support for automatic device management, data movement, and code generation. 
This is possible by exploiting the design of VM-based languages: Tornado simply augments the underlying VM with support for OpenCL by using the JVMCI (Java Virtual Machine Compiler Interface); similarly, to Indigo.
The JVMCI allows efficient access to low-level information inside the JVM, such as a methods bytecodes and profiling information.
Using this information Tornado is able to JIT compile Java bytecode to execute on OpenCL compatible devices.

\begin{figure}
\centering
\includegraphics[width=\columnwidth]{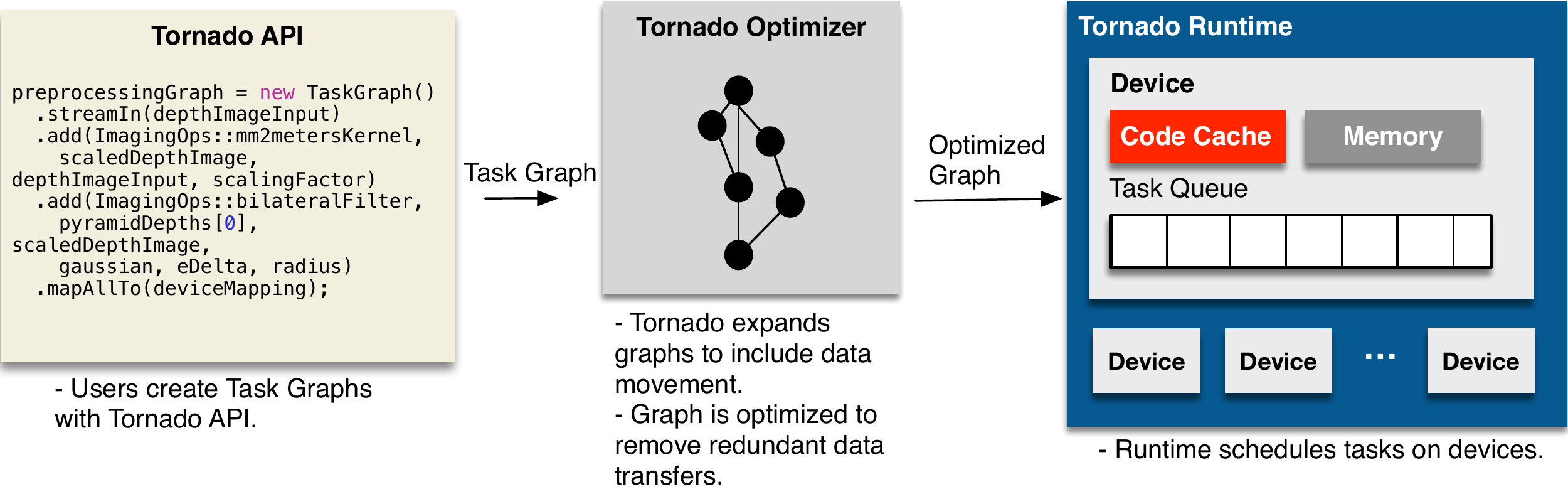}
\caption{Tornado outline.}
\label{fig:tornado-outline}
\end{figure}

As depicted in Figure \ref{fig:tornado-outline}, the Tornado API provides developers with a task-based programming model.
In Tornado, a task can be thought of as being analogous to a single OpenCL kernel execution.
This means that a task must encapsulate the code it needs to execute, the data it should operate on, and some meta-data.
The meta-data can contain information such as the device it should execute on or profiling information.
The mapping between tasks and devices is done at a task-level granularity; meaning each task is capable of being executed on a different piece of hardware.
These mappings can be provided either by the developer or by the Tornado runtime; the mappings are dynamic and have the ability to change anytime.

Instead of focusing on scheduling individual tasks, Tornado allows developers to combine multiple tasks together to form a larger schedulable unit of work (called a \textit{task-graph}).
This approach has a number of benefits: firstly, it provides a clean separation between the code which co-ordinates tasks execution and the code which performs the actual computation; and secondly, it allows the Tornado runtime system to exploit a wider range of runtime optimizations.
For instance, the task-graph provides the runtime system with enough information to determine the data dependencies between tasks.
By using this knowledge, the runtime system is able to exploit any available task parallelism by overlapping the execution of task execution and data movement.
It also provides the runtime system with the ability to eliminate any unnecessary data transfers that would occur because of read-after-write data dependencies between tasks.

To increase developer productivity, Tornado is designed to make offloading computation as transparent as possible.
This is achieved via its runtime system which is able to automatically schedule data transfers between devices and handle the asynchronous execution of tasks.
Moreover, the JIT compiler provides support for user-guided parallelization.
The result is that developers are able to rapidly develop portable heterogeneous applications which can exploit any OpenCL compatible device in the system.

\subsection{Binary Instrumentation Layer}
\label{bininst}
Beehive integrates a number of binary instrumentation tools to enable research and rapid prototyping of novel micro-architectures and ISA extensions.
Along with the well-established Intel's PIN tool \cite{PIN}, Beehive integrates the newly introduced MAMBO \cite{Gorgovan:2016:MLD:2899032.2896451}, and MAMBO-x64 \cite{D'antras:2016:OIB:2899032.2866573} tools for ARMv7 and AArch64 architectures.

\subsubsection{MAMBO}~

\noindent MAMBO is a low-overhead dynamic binary instrumentation and modification tool for the ARM architecture which currently supports ARMv7 and the AArch32 execution state of ARMv8.
In the context of Beehive, the initial performance of MAMBO has been further improved since its first release.
The introduced optimizations include: 
\begin{itemize}
	\item A novel scheme to enable hardware return address prediction for dynamic binary translation.
	\item A novel software indirect branch prediction scheme for polymorphic indirect branches.
	\item A number of micro-architectural specific optimizations such as usage of huge pages for internal data.
\end{itemize}

While the initial version of MAMBO achieves a geometric mean overhead of 28\% on a Cortex-A9 (a dual-issue out-of-order superscalar
processor with 8 to 11 pipeline stages) and of 34\% on a Cortex-A15 (a triple-issue out-of-order superscalar processor with 15 to 24 pipeline stages), 
the introduced optimizations reduce the overhead on the two systems to 15\% and 21\% respectively.

\subsubsection{MAMBO-X64}~

\noindent The introduced ARM AArch64 architecture is a 64-bit execution mode with a new instruction set which retains binary compatibility with ARMv7 32-bit execution mode.
Due to the need to support the large number of existing 32-bit ARM applications, current implementations of AArch64 processors include hardware support for ARMv7. 
However, this support comes at a cost in hardware complexity, power usage, and verification time.

MAMBO-X64 is a dynamic binary translator which executes 32-bit ARM binaries (both single-threaded and multi-threaded) using the AArch64 instruction set. 
The integration of MAMBO-X64 into Beehive creates a path for experimentation for future processors to drop hardware support for the
legacy 32-bit instruction set while retaining the ability to run ARMv7 applications.

In the context of Beehive, the performance of MAMBO-X64 has been further improved by employing a number of novel optimizations such as:
mapping ARMv7 floating-point registers to AArch64 registers dynamically, generating traces that harness hardware return address prediction, and
efficiently handling operating system signals.
After applying the aforementioned optimizations, on SPEC CPU2006 \cite{speccpu}, we measured a very low geometric mean average
performance overhead of 0.2\%, 3.3\% and 8.3\% on X-Gene, Cortex-A57, and Cortex-A53 processors respectively. 
The performance of MAMBO-X64 also scales to multi-threaded applications, with an overhead on the
PARSEC \cite{bienia11benchmarking} multithreaded benchmark suite of only 2.1\% with 1, 2 and 4 threads, and 4.9\% with 8 threads.

\subsection{Hardware/FPGA Layer}
\label{mast}
As depicted in Figure \ref{beehive_stack}, Project Beehive targets a variety of hardware platforms and therefore significant effort is being placed in providing the appropriate support for the compilers and runtimes of choice.
Besides targeting conventional CPU/GPU systems, it is also possible to target FPGA systems such as the Xilinx Zynq ARM/FPGA System on Chip (SoC). 

In order to efficiently program FPGAs from high level programming languages, we developed MAST: a Modular Acceleration and Simulation Technology.
MAST consists of a hardware/software library and tools allowing the rapid development of systems using ARM based FPGAs.  
From the hardware perspective it consists of a standardized interface which allows IP blocks to be identified and locked for use by processes running on the ARM processor. 
All IP blocks feature an AXI slave port, used for configuration and low speed communication, and optionally an AXI master port to provide high speed access to the system memory of the ARM processor, typically via the ACP port to provide cache coherency. 
Currently hardware design is carried out using Bluespec System Verilog \cite{Arvind:2003:BLH:823453.823860}, with interface modules conforming to the hardware. 
The software library, which is entirely in user space, provides a hardware manager which can be used to discover IP on the programmable logic and allocate it a specific process thread.
The software library also provides a simple interface with IP blocks between the virtual memory world of the processor and the physical memory required by the hardware, where either the library or the host application can perform memory allocation.

\subsection{Simulation Layer}
\label{simul}
Besides running directly on real hardware, Beehive offers the opportunity to conduct micro-architectural research via its advanced simulation infrastructure. 
The two simulators of choice, with diverse characteristics, ported to the Beehive platform are: Gem5 \cite{gem5} and ZSim \cite{ZSim}. 
While Zsim offers a fast and high accurate simulation time on x86 ($\approx$ 10 MIPS in our experiments), Gem5 provides a slower yet more detailed full-system simulation framework for numerous architectures.

\subsubsection{Gem5}~
\label{gem5}

\noindent The Gem5 full-system simulator has been adopted and augmented in the following ways:
\begin{itemize}
	\item \textbf{Integration with other architectural simulators:} 
	A new interface layer has been developed within the Gem5 full-system simulator \cite{Binkert:2011:GS:2024716.2024718} to facilitate easy integration with a range of architectural simulators as depicted in Figure \ref{fig:gem5}. 
	
	\begin{figure}
	\centering
	\includegraphics[width=\columnwidth]{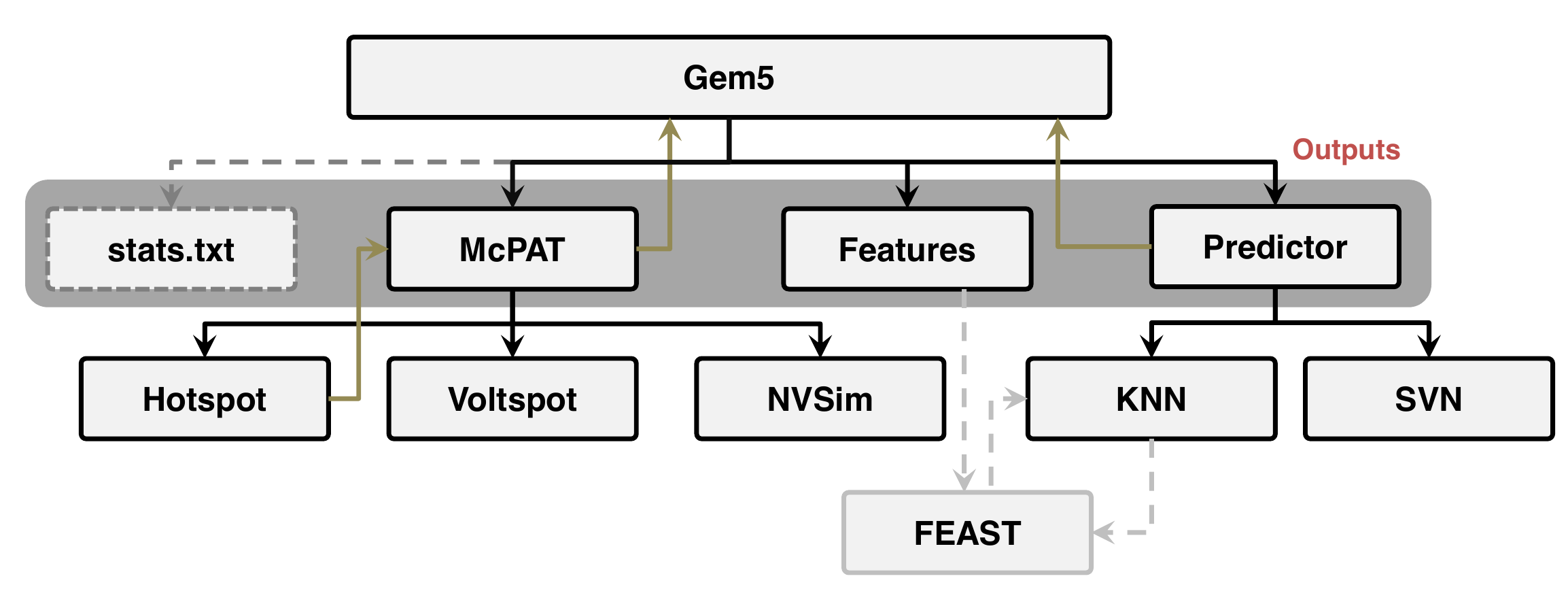}
	\caption{Beehive's Gem5 stack.}
	\label{fig:gem5}
	\end{figure}
	
	The statistics package has been augmented to allow statistics to be assigned to groups, specified at run-time and manipulated (output and reset) independently, without affecting the total values of the statistics or requiring updates to the code base. 
	This allows new architectural simulators to be invoked from within the Gem5 simulator by using standard C++ template code. 
	Current simulators integrated into the Gem5 framework include:
	
	1) McPAT \cite{5375438} and Hotspot \cite{1650228}: The power and temperature modelers provided by those tools are conjoined to provide accurate temperature-based leakage models. 
	Power samples may be triggered from within the Gem5 simulator, at intervals between 10ns to 10us (allowing transient traces to be generated for benchmarks), and from within the 	simulated OS (allowing accurate power and temperature figures to be used within user space programs). 
	There is around a 10\% simulation time overhead for temperature and power modelling with 10us samples.
	
	2) Voltspot \cite{6853199}: In order to measure Voltage noise events caused by power-gating or switching patterns in Multicore SOCs over realistic workloads, the Voltspot 	simulator has been incorporated into the framework.
	The additional statistics generated allow nanosecond timing of events to be recorded while using samples of courser granularity.

	3) NVSim \cite{dong2014nvsim}: The non-volatile memory simulator NVSim has been incorporated into the simulation infrastructure. 
	NVSim can be invoked by McPat (alongside the conventional SRAM modeling tool Cacti \cite{li2011cacti}) allowing accurate delay, power, and temperature modeling of non-volatile 	memory anywhere in the memory hierarchy.
	\item \textbf{Machine Learning and Data Analytics techniques:}
	The interface layer has also been used to allow machine-learning/data-analytics techniques to be incorporated within the simulation framework.
	Machine-learning techniques are used to analyze statistical patterns in the data aiding in the creation of hardware-predictors for power-management, prefetching, branch-prediction etc.
	The statistics package allows for the specification of \textit{features} at runtime.
	Features are defined as a statistic over a given period (e.g. the branch mispredict rate over 1us, or the L2 cache miss-rate over 10ms).
	Features are specified at run-time and can be accessed periodically or triggered from events within the simulator and the statistics package guarantees to return the features over their specified time (within an error range which is also set at run-time).
	The FEAST toolkit \cite{brown2012conditional} has been incorporated into the framework (Figure \ref{fig:gem5}) to allow for (offline) feature selection.
	Packages for online K-nearest neighbour (KNN) and Support Vector Machine regression have been incorporated into the framework to allow for online prediction once the features have been chosen.
	Interaction between the simulator and the predictors is controlled by the statistics package again allowing for the prediction to be triggered within the Gem5 simulator code or from within the simulated OS.
	\item \textbf{Resiliency and Fault-Injection:} A critical aspect of any computer system is its dependability evaluation \cite{Li:2008:UPH:1353535.1346315, 7314163, 1311888}. 
	The accurate identification of vulnerabilities assists computer architects to carefully plan for low cost and high energy efficient resiliency mechanisms. 
	On the contrary, inaccurate dependability assessment often results on over-designed microprocessors impacting negatively time-to-market and product costs.
	To aid dependability studies, we developed a fault injection framework that adheres to the following principles: 1) Flexibility: easy to setup, define and perform fault injection experiments, 2) Reproducibility: enable reproducible experiments, 3) Generality: support a wide set of ISAs in a uniform way performing comparative studies, and 4) Scalability: easily deployed to multi-core designs.
	
	\begin{figure}
	\centering
	\includegraphics[scale=0.3]{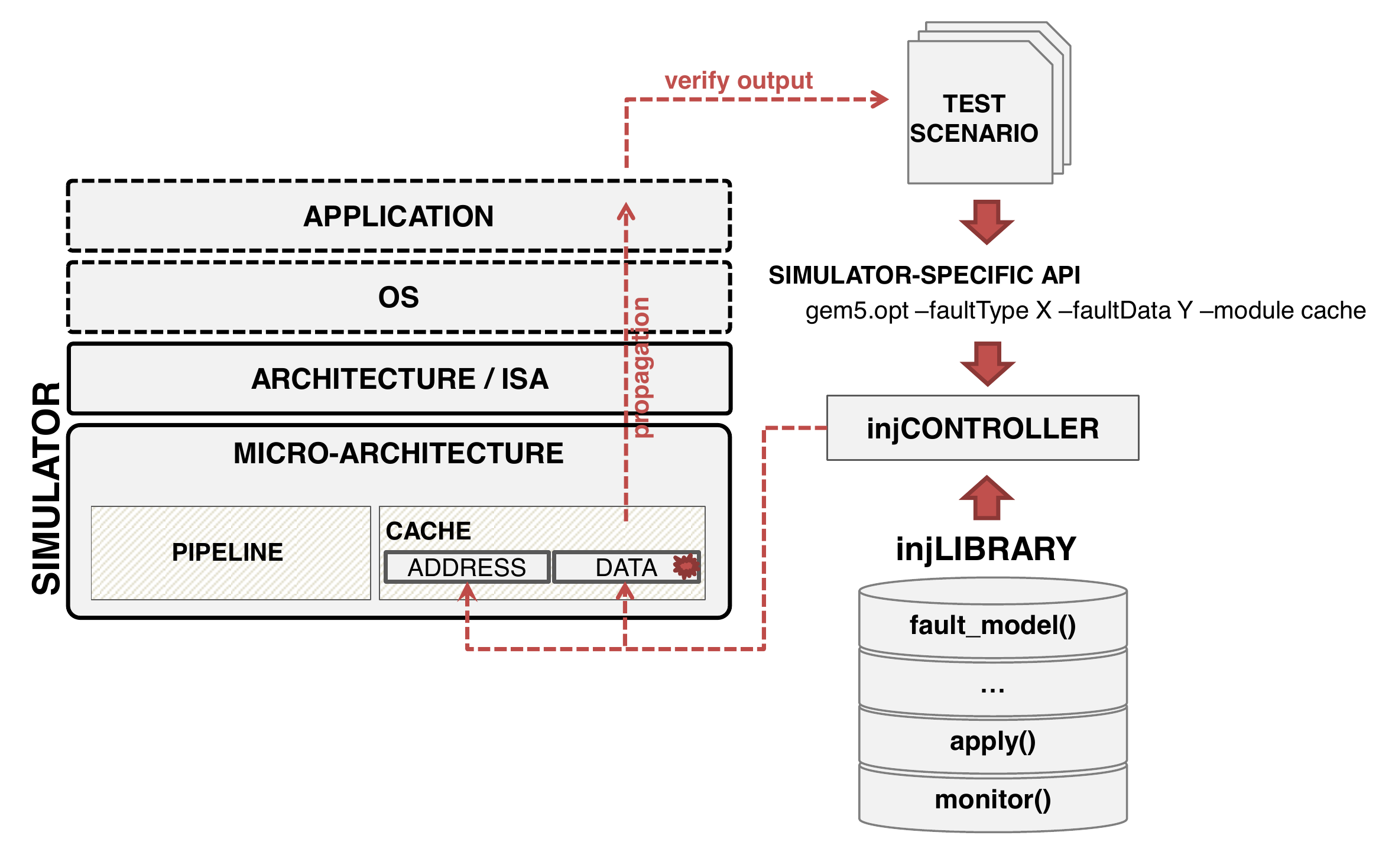}
	\caption{Beehive's fault injection tool.}
	\label{fig:error}
	\end{figure}
	Figure \ref{fig:error} depicts the floor-plan of the fault injection tool.
	The developed fault injection framework is built on top of Gem5 and operates as follows: A user-defined test scenario is translated into a set of fault injection arguments using a 	simulator-specific API. 
	The injection library implements all the necessary simulation calls: 
	(i) fault\_model(): setup of a transient, intermittent or permanent fault model \cite{Biswas:2005:CAV:1080695.1070014, 1225959, 5432157}. 
	Transient faults are modeled by flipping the value of a randomly selected bit in a randomly selected time window within simulation. 
	Intermittent faults are modelled by setting the state of storage elements to one (stuck-at-1) or zero (stuck-at-0), in a randomly selected time window, for a random period. 	Moreover, permanent faults set the state of storage element persistently to one or to zero. 
	Finally, multi-bit fault injections, having a combination of the aforementioned models, are also supported. 
	(ii) apply(): injects the faults into a user-defined location (e.g. L1, L2 cache, etc.);
	and (iii) monitor(): logs and clusters the fault injection output. 
	Finally, the injection controller, the kernel of the framework, communicates with the injection library and orchestrates the actual fault injection based on the user-defined 	arguments.
\end{itemize}

\subsubsection{ZSim}~
\label{maxinezsim}

\noindent The ZSim simulator, a user-level x86-64 simulator with an OOO-core model of the Westmere (Nehalem) micro-architecture, has been augmented in order to run managed workloads on MaxineVM resulting in the MaxSim simulation platform \cite{rodchenko2017maxsim}.
Alternative options such as the Sniper \cite{Sniper} simulator that runs with JikesRVM ~\cite{JikesOnSniper}, or the full-system Gem5 simulator were considered but abandoned due to a number of limitations: Sniper can only run in a 32-bit mode, while Gem5 has a relatively low simulation speed.
Finally, in order to perform energy and power estimations, we integrated the McPAT \cite{McPAT} tool into the ZSim simulator following the methodology proposed by the Sniper simulator \cite{Sniper-McPAT}. 
The methodology necessitated the implementation of a number of extra micro-architectural events in ZSim such as the number of predicted branches and floating point micro-operations.

%% file: usecase.tex
\section{SLAM Applications}
\label{usecase}
\subsection{KinectFusion}
\begin{figure}[h]
\centering
	\includegraphics[width=\columnwidth]{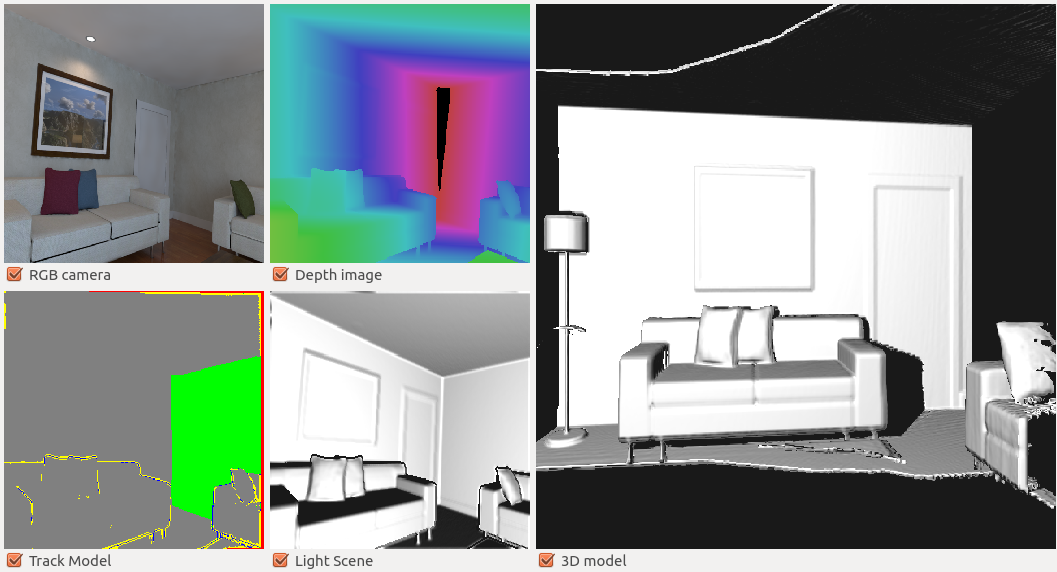}
	\caption{RGB-D camera combines RGB with Depth information (top left and middle). The tracking (left) results in the 3D reconstruction
of the scene (right).}
	\label{fig:slambench}
\end{figure}

\begin{figure}[t]
\centering
	\includegraphics[scale=0.3]{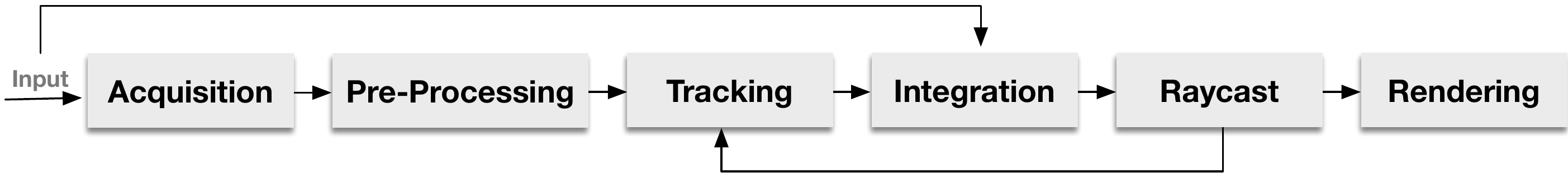}
	\caption{KinectFusion Pipeline.}
	\label{fig:kfusion_pipeline}
\end{figure}

To showcase the capabilities of the ZZZ platform, we focused on emerging applications which are becoming significant both in desktop and mobile domains: real-time 3D scene understanding in Computer Vision.   
In particular, we investigate SLAMBench a complex Simultaneous Localization and Mapping (SLAM) application which implements the KinectFusion (KFusion) algorithm. 
SLAM applications are challenging due to the amount of computation needed per frame and the programming complexity of achieving high performing implementations.
SLAMBench allows the reconstruction of a three-dimensional representation from a stream of depth images produced by a RGB-D camera (Figure \ref{fig:slambench}), such as the Microsoft Kinect.
Typically, the slower the frames are processed, the harder it is to build an accurate model of the scene.
Each of the depth images is used as input to the six-stage processing pipeline shown in Figure \ref{fig:kfusion_pipeline}:
\begin{itemize}
	\item \textbf{Acquisition} obtains the next RGB-D frame; either from a camera or from a file.
	\item \textbf{Pre-processing} is responsible for cleaning up the incoming data using a bilateral filter and standardizes the units used for measurement.
	\item \textbf{Tracking} estimates the new pose of the camera; it builds a point cloud from the current data frame and matches it against a reference point cloud, produced from the raycasting step, using an iterative closest point (ICP) algorithm.
	\item \textbf{Integrate} fuses the current frame into the internal model, if a new pose has been estimated.
	\item \textbf{Raycast} using raycasting the pipeline can construct a new reference point cloud from the internal representation of the scene.
	\item \textbf{Rendering} this stage uses the same raycasting technique to visualize the 3D scene.
\end{itemize}

It should be noted that the pipeline has a feedback loop.
Each of the pipeline stages is composed from a number of different kernels.
In the original KinectFusion implementation, a kernel represents a separate region of code which is executed on the GPU. 
In a typical pipeline execution KinectFusion will execute between 18 and 54 kernels (best and worst case scenarios). 
The variation is dependent on the performance of the ICP algorithm, if it is able to estimate the new camera pose quickly then less kernels will be executed. 
This means that to achieve a real-time performance of 30 frames per second, the application will need to sustain the execution of between 540 and 1620 kernels every second.

\subsection{Programmability Vs. Performance} 
\label{opts}
SLAMBench offers baseline and high-performing implementations of KinectFusion in C++, OpenMP, CUDA, and OpenCL.
In order to achieve the QoS targets of Computer Vision (typically over 30 FPS), KinectFusion has to be heavily parallelized on GPGPUs and therefore the CUDA and OpenCL implementations are those matching the required targets.
Developing on CUDA or OpenCL, however, comes with a number of drawbacks.
The first one is code complexity and productivity while the second one is portability since applications have to recompiled and tuned for each target hardware platform.

To tackle the aforementioned problems and to showcase the capabilities of ZZZ, we decided to experiment with Computer Vision applications in Java; a language that up-to-now was not considered for such high performing and demanding applications.
Implementing SLAMBench, and in general Computer Vision applications, in Java provides a trade-off between programmability efforts and performance.

While Java can provide rapid prototyping, in contrast to writing OpenCL or CUDA, vanilla and un-optimized implementations can not meet the QoS requirements.
We use the Java programming language as a challenge in order to build and optimize Computer Vision applications aiming to achieve real-time 3D space reconstruction.
After having developed and validated a serial implementation of SLAMBench, we performed a performance analysis and identified performance bottlenecks.
Then, we utilized ZZZ to apply a number of co-designed acceleration and optimization techniques to the various stages of SLAMBench.
The acceleration techniques span from custom FPGA acceleration of certain kernels to full-application acceleration through co-designed object compaction and GPGPU off-loading.

%% file: results.tex
\section{Evaluation}
\label{results}
The following subsections describe the acceleration and optimizations techniques applied to SLAMBench via the Beehive platform along with the experimental results.
The hardware and software configurations for each optimization are presented in Table \ref{platform}.

\newcolumntype{P}[1]{>{\centering\arraybackslash}p{#1}}
\begin{table*}[]
\centering
\small
\begin{tabular}{|c|P{1cm}|P{1cm}|P{1cm}|P{1cm}|P{1cm}|P{1cm}|P{5cm}|P{5cm}|}
\hline
\textbf{Optimization} & \multicolumn{6}{c|}{\textbf{1: GPU Acceleration}} & \textbf{2: FPGA Acceleration} & \textbf{3: HW/SW Co-Designed Object Compaction} \\ \hline
\textbf{Beehive Module} & \multicolumn{6}{c|}{OpenJDK, Graal, Tornado} & OpenJDK, Maxine, MAST & Maxine, Zsim, McPAT \\ \hline
\multicolumn{9}{|c|}{\textbf{Hardware}} \\ \hline
\textbf{CPU} & \multicolumn{6}{P{4cm}|}{Intel Xeon E5-2620 @ 2Ghz} & Xilinx Zynq 706 board, ARMv7 Cortex A9 @ 667Mhz & Simulated: x86-64 Nehalem @ 2.64Ghz \\ \hline
\textbf{Cores} & \multicolumn{6}{P{4cm}|}{12 (24 Threads)} & 2 & 4 \\ \hline
\textbf{L1} & \multicolumn{6}{P{4cm}|}{32KB per core, 8-way} & 32KB per core & 32KB, 8-way, LRU, 4 cycles \\ \hline
\textbf{L2} & \multicolumn{6}{P{4cm}|}{256KB per core, 8-way} & 512KB per core & 256KB, 8-way, LRU, 6 cycles \\ \hline
\textbf{L3} & \multicolumn{6}{P{4cm}|}{15MB, 20-way} & - & 8MB, 16-way, hashed, 30 cycles \\ \hline
\textbf{RAM} & \multicolumn{6}{P{4cm}|}{32GB} & 1GB & 3GB, DDR3-1066, 47 cycles \\ \hline
\textbf{GPU} & \multicolumn{6}{P{4cm}|}{NVIDIA Tesla K20m @ 0.705Ghz, OpenCL 1.2} & - & - \\ \hline
\textbf{Extensions} & \multicolumn{6}{c|}{-} & MAST FPGA & AGU Extensions \\ \hline
\multicolumn{9}{|c|}{\textbf{Software}} \\ \hline
\textbf{JVM} & \multicolumn{6}{c|}{OpenJDK, Graal} & Maxine ARMv7, OpenJDK\_1.7.0\_40 & Maxine x86 \\ \hline
\textbf{OS} & \multicolumn{6}{c|}{CentOS 6.8 (Kernel 2.6.32)} & Linux 3.12.0-xilinx-dirty & Ubuntu 14 LTS 3.13.0-85 \\ \hline
\end{tabular}
\caption{Beehive, Hardware, and Software experimental configurations.}
\label{platform}
\end{table*}

\subsection{GPU Acceleration}
\begin{figure}
\centering
\includegraphics[width=\columnwidth]{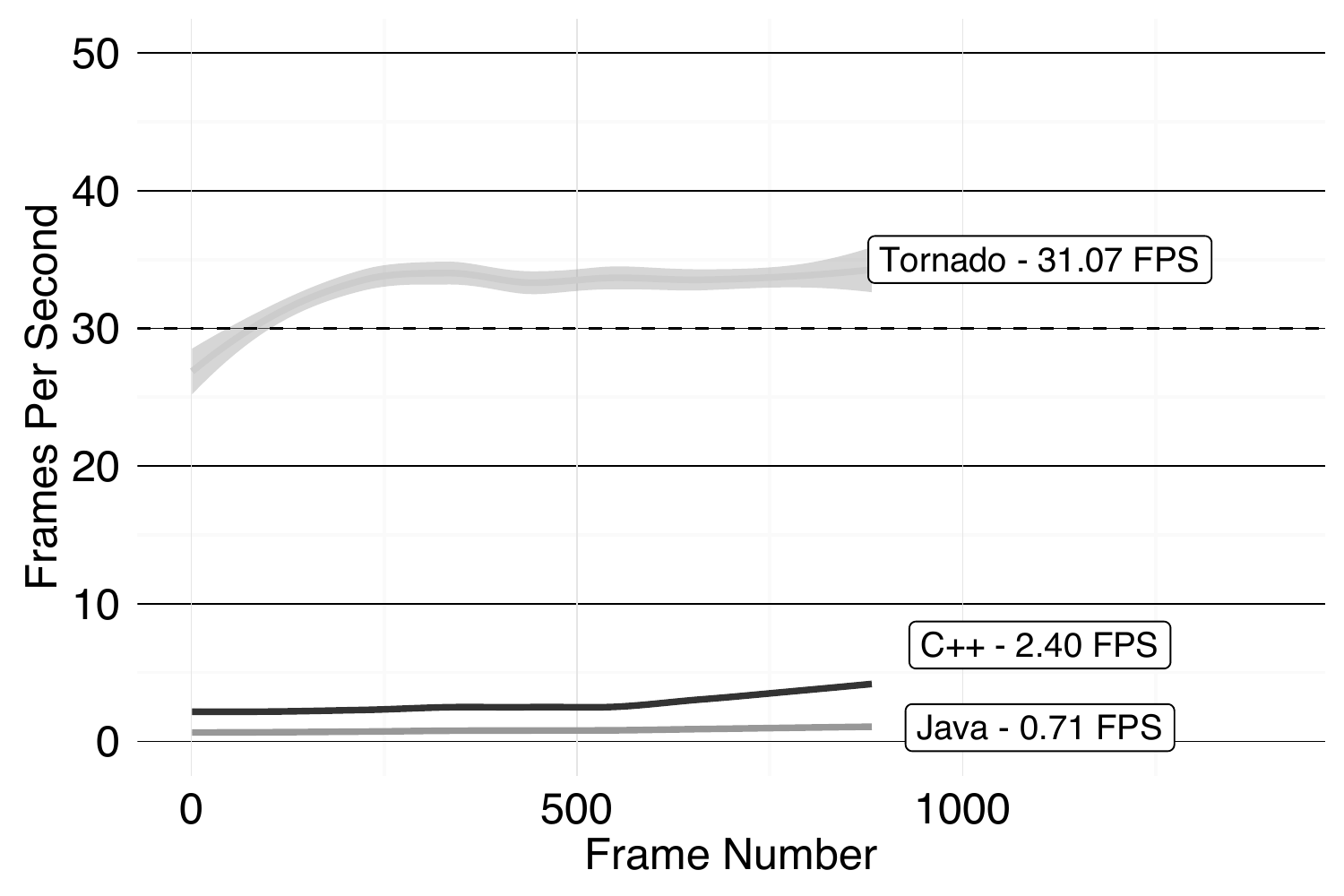}
\caption{FPS achieved of Tornado versus baseline Java and C++ implementations.}
\label{fig:tornado-fps}
\end{figure}

\begin{figure}
\centering
\includegraphics[width=\columnwidth]{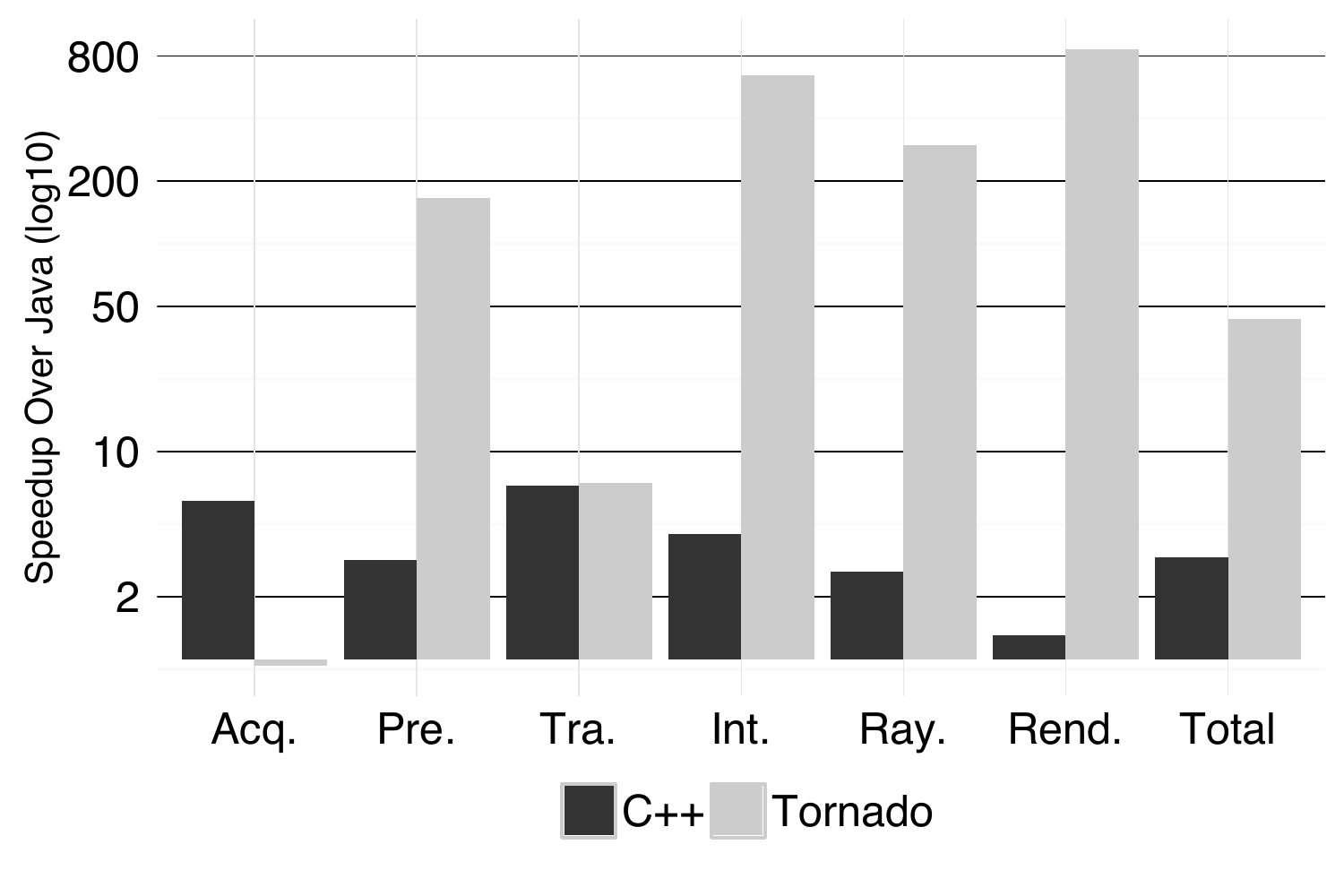}
\caption{Tornado Speedup versus serial Java and C++ implementations per KFusion stage.}
\label{fig:tornado-speedup}
\end{figure}

GPU acceleration has been applied to SLAMBench through Tornado (Section \ref{tornado}).
All kernels but one\footnote{Acquisition can not be accelerated because the input is serially obtained from a camera or a file.} of KinectFusion have been dynamically compiled and offloaded for GPGPU execution through OpenCL code emission.
Figures \ref{fig:tornado-fps} and \ref{fig:tornado-speedup}, illustrate the performance and speedup of the accelerated KinectFusion version respectively.

As depicted in Figure \ref{fig:tornado-fps}, the original validated version of KinectFusion can not meet the QoS target of real-time Computer Vision applications (0.71 FPS on average).
Both the serial versions of Java and C++ perform under 3 FPS with the C++ version being 3.3x faster than Java.
By accelerating KinectFusion through GPGPU execution we manage to achieve a constant rate of over 30 FPS (31.07 FPS) across all frames (802) from the ICL-NUIM dataset \cite{2014Handa} (Room 2 configuration).
In order to achieve 30 FPS, all kernels have been accelerated by up to 861.26x with an average of 43.37x across the whole application, as depicted in Figure \ref{fig:tornado-speedup}.
By utilizing Beehive and its GPU acceleration infrastructure, we manage to accelerate a simple un-optimized serial Java version of a KinectFusion algorithm meeting its QoS requirements in a transparent to the developer manner.

\subsection{FPGA Acceleration}
FPGA acceleration has been applied to SLAMBench through the MAST acceleration functionality of Beehive (Section \ref{mast}).
In the context of our initial investigation into FPGA acceleration we have selected the \texttt{pre-processing} stage that contains
two computational kernels that: i) scale the depth camera image from \texttt{mm} to \texttt{meters}, and ii) apply a bilateral filter to produce a filtered scaled image. 
A filter is applied to the scaled image in order to reduce the effects of noise in depth camera measurements.
This includes missing or invalid values due to the characteristics of the 3D space\footnote{For example, null or invalid measurements are obtained when the surfaces are translucent, and/or the angle of incidence of the infrared radiation from the depth camera is too acute to be reflected back to the camera's sensors.}.

In order to improve the execution time in Java, we merged the two routines into a single routine reducing the streaming of data to and from the FPGA device. 
The offloading to the FPGA is accomplished by using the Java Native Interface (JNI) mechanism to interface with our MAST module (Section \ref{mast}).
The JNI stub extracts C-arrays of floating point values from the Java environment that represent the current input raw depth image from the camera, and the current output scaled filtered image.
The JNI stub, in turn, converts the current raw depth image into an array of short integers which is memory allocated (through \texttt{malloc}) on first execution of the JNI stub. 
The FPGA hardware environment is also initialized during first execution, and consequently the hardware performs the merged scaling and filtering operation.
Subsequent executions only need to perform a call to extract C-arrays and to, finally, release the output scaled and filtered image array back to the Java environment.

As depicted in Table \ref{tab:fpga}, FPGA acceleration improves performance by 43x and 22x on MaxineVM and OpenJDK respectively.
The difference in both execution times and speedups from both VMs stem from the fact that OpenJDK produces more optimal code than MaxineVM (Section \ref{Runtime}).

\begin{table}[]
\centering
\small
\begin{tabular}{|c|c|c|c|}
\hline
%\noalign{\smallskip}
\textbf{VM} & \textbf{No FPGA }		 & \textbf{With FPGA} & \textbf{Speedup}\\
\textbf     & \textbf{Acceleration}  & \textbf{Acceleration} & \\
%\noalign{\smallskip}
\hline
%\noalign{\smallskip}
Maxine VM  			 & 2.20 & 0.05 & 43x \\
OpenJDK          	 &   0.66 &  0.03 & 22x \\
%\noalign{\smallskip}
\hline
\end{tabular}
\caption{Performance and speedup of KFusion's pre-processing stage with and without FPGA acceleration (mean execution time, in seconds, over 78 frames).}
\label{tab:fpga}
\end{table}

\subsection{HW/SW Co-Designed Object Compaction}
This generic optimization applies to all Java objects and regards class information elimination from object headers.
This is achieved by utilizing tagged pointers; a feature currently supported by ARM AArch64 \cite{ProgrammersGuideForARMv8} and SPARC M7 \cite{M7NextGenerationSPARC}.
In order to apply that optimization, changes both at the Virtual Machine and at the hardware layers have to be performed.
In our case, it has been applied to SLAMBench through the Maxine/ZSim stack \cite{rodchenko2017OHE} (Section \ref{maxinezsim}).

\begin{figure}
  \includegraphics[width=\columnwidth]{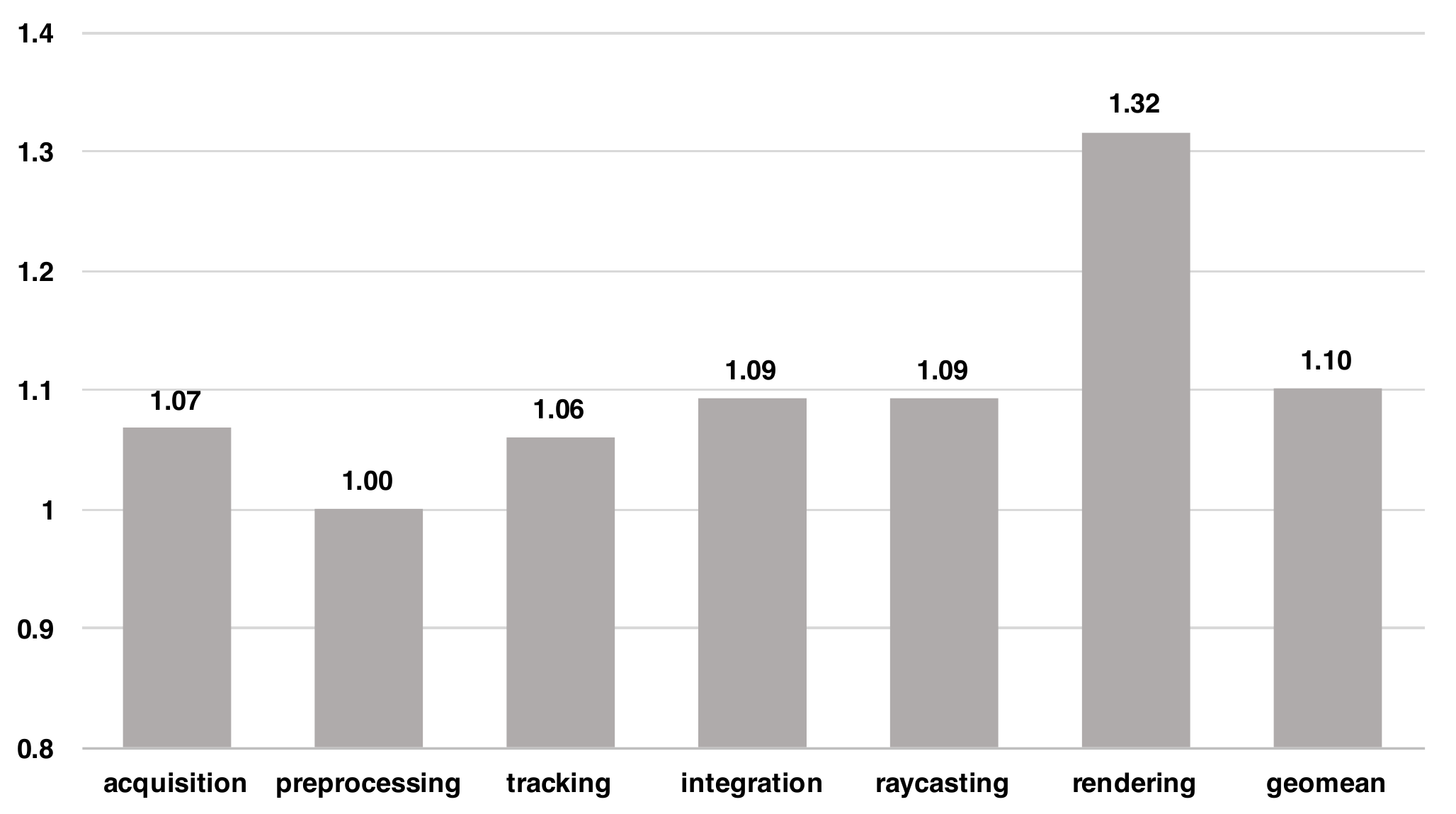}
  \caption{Performance improvements of class information elimination in SLAMBench.}
  \label{fig:kfusion_ohe_speedup}
  \vspace{1mm}
\includegraphics[width=\columnwidth]{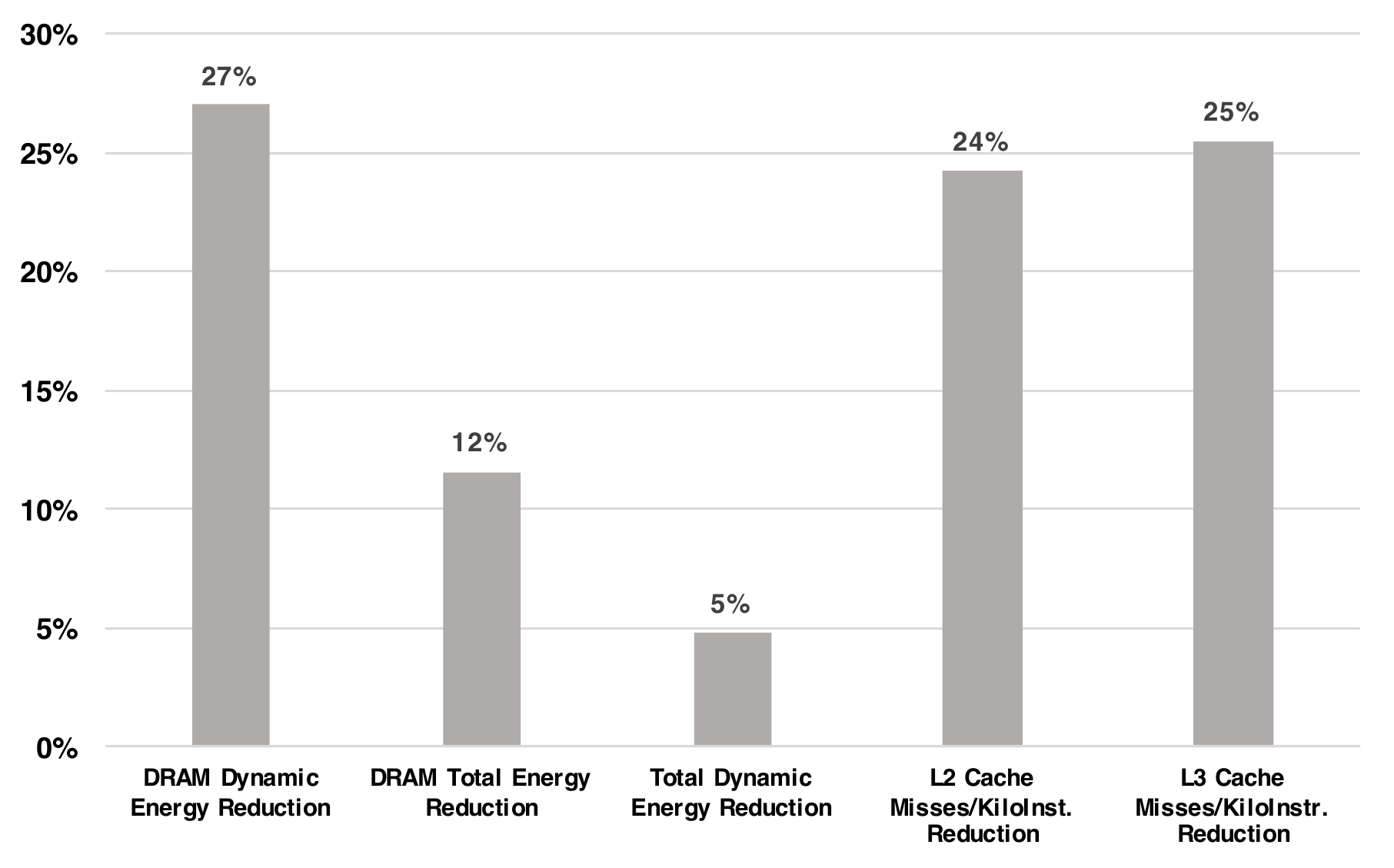}
\caption{Energy and Cache Miss improvements of class information elimination in SLAMBench.}
\label{fig:kfusion_ohe_microarch}
\end{figure}

Object-oriented programming languages have the fundamental property of associating type information with objects allowing them to perform various tasks such as virtual dispatch,
introspection, and reflection. 
Typically, this is performed by maintaining an extra pointer per object to its associated type information.
To save that extra heap space per object, we utilize tagged pointers in order to encode class information inside object addresses.
By extending ZSim to support tagged pointers in x86 and by extending the Address Generation Unit (AGU) at the micro-architectural level we managed to expose tagged addresses at the JVM level.
Instead of maintaining the extra pointer per object, we exploit the unused bits of tagged pointers to encode that information.
The proposed optimization, which is orthogonal to any application running on top of the JVM, has been applied to SLAMBench and results are shown in Figures \ref{fig:kfusion_ohe_speedup} and \ref{fig:kfusion_ohe_microarch}.

\captionsetup[figure]{skip=0pt}

As depicted in Figure \ref{fig:kfusion_ohe_speedup}, by employing the co-designed optimization for eliminating class information from object headers we managed to achieve up to 1.32x speedup with an average of 1.10x across all stages of SLAMBench.
Furthermore, as depicted in Figure \ref{fig:kfusion_ohe_microarch} the optimization resulted in up to 27\% Dynamic DRAM energy, 12\% total DRAM energy, and 5\% total dynamic energy reductions.
The energy reductions correlate with improvements in cache utilization of 24\% and 25\% in L2 and L3 caches respectively.
The observed benefits of employing the introduced optimization derive from the fact that by compressing object sizes by one word we managed to: 1) improve cache utilization, 2) reduce garbage collection invocations (from 10 to 7) due to heap savings, and 3) improve retrieval time for class information due to the introduced minimal hardware extension.

%% file: related_work.tex
\section{Related Work}
\label{related}
Although heterogeneity is the dominant design approach, its programming environment is extremely challenging. 
Delite \cite{6113791, Chafi:2011:DAH:1941553.1941561} is a compiler and runtime framework for parallel embedded domain-specific languages \cite{Sujeeth:2013:FGH:2517208.2517220, Sujeeth11optiml:an}.
Its goal is to facilitate heterogeneous programming to efficiently exploit the underlying heterogeneous hardware capabilities.
SWAT \cite{Grossman:2016:SPI:2907294.2907307} is a software platform that enables native execution of Spark applications on heterogeneous hardware. 
Furthermore, OpenPiton \cite{Balkind:2016:OOS:2872362.2872414} is an open source many-core research framework covering only the hardware layer, X-Mem \cite{GottschoGSSG16} is an open-source software tool that characterizes the memory hierarchy for cloud computing, and Minerva \cite{Minerva} is a HW/SW co-designed framework for deep neural networks. 
In contrast to the aforementioned approaches, the Beehive framework is a hardware/software experimentation platform that enables co-designed optimizations for runtime and architectural research.
covering all applications and compute stack.
Regarding GPGPU Java acceleration, a number of approaches such as APARAPI \cite{amd;;aparapi}, Ishizaki et. al. \cite{ishizaki;2015;compiling-and-o}, Rootbeer \cite{pratt-szeliga;2012;rootbeer:-seaml}, and Habanero-Java \cite{hayashi;2013;accelerating-ha}, exist.
Beehive's Tornado module differs due to its dynamic nature and its co-operation with other parts of the framework such as MAST.

%% file: conclusions.tex
\section{Conclusions and Future Work}
\label{conclusions}
In this paper, we introduced Beehive: a hardware/software co-designed platform for full-system runtime and architectural research. 
Beehive builds on top of existing state-of-the-art as well as novel components at all layers of the platform.
By utilizing Beehive, we managed to accelerate a complex Computer Vision application in three distinct ways: GPGPU acceleration, FPGA acceleration, and by compacting objects
in a hardware/software co-designed manner.
The experimental results proved that we managed to achieve real-time 3D space reconstruction ($>$30 fps) of the KFusion application, after accelerating it by up to 43$\times$.

Our vision regarding Beehive is to improve both its integration and performance throughout all the layers.
In the long term, we aim to unify the platform's components under a semantically aware runtime increasing developer productivity.
Furthermore, we plan to define a hybrid ISA between emulated and hardware capabilities.
This ISA will provide a roadmap of movement of interactions between abstractions offered in software and in hardware. 
Finally, we plan to work on new hardware services for scale out and representation of volatile and non-volatile communication services.
This will provide a consistent view of platform capabilities across heterogeneous processors for Big Data and HPC applications.